\documentclass[aps,prb,amsmath,twocolumn,amssymb,titlepage]{revtex4-1}

\usepackage{graphicx}
\usepackage{nicefrac}
\usepackage{amsfonts}

\usepackage{amssymb}
\usepackage{amsmath} 

\usepackage{subfigure}
\usepackage{multirow} 
\usepackage{tabularx} 
\usepackage{array}
\usepackage{marvosym}
\usepackage{units}

\usepackage{tensor} 
\usepackage{braket}

\usepackage{bm}
\usepackage{hyperref}

\begin{document}
	\title{Analysis to closed surface-wave photonic crystal waveguides based on coupled-resonator optical waveguide theory}
	\author{Yonghui Zheng(1,2)}
	\author{Chang Wang(1,2)}
	\thanks{Footnote.}
	\email{cwang@mail.sim.ac.cn}
	\author{JunCheng Cao(1,2)}
	\thanks{Footnote.}
	\email{jccao@mail.sim.ac.cn}
	
	\affiliation{1.Key Laboratory of Terahertz Solid State Technology, Shanghai Institute of Microsystem and Information Technology, Chinese Academy of Sciences, 865 Changning road, Shanghai 200050, China.}
	\affiliation{2.Center of Materials Science and Optoelectronics Engineering, University of Chinese Academy of Sciences, Beijing 100049, China}

	
\begin{abstract}
	Traditionally, one can construct a waveguide by introduce defects into surface-wave photonic crystals (SPCs). Here we propose a new structure named closed SPC that can introduce waveguide modes out of photonic bandgap of surface-wave photonic crystal. In this paper, we have comprehensively analyzed dispersion relation, group velocity, normalized transmission and electric field distribution of closed SPC waveguides, and propose several methods to improve performance of the waveguide based on coupled-resonator optical waveguide theory. These methods can improve transmission efficiency from $10\%$ to $60\%$ and eliminate intraband oscillation by adjusting eigenfrequency $\Omega$ and coupling factor $\kappa_1$. These methods are applicable to both single mode and multi mode situations. This letter also paves a new way for improving the performance of other coupled-resonator waveguides.
\end{abstract}

	\maketitle
	
	\section{Introduction}
	Waveguide, a basic component, plays a vital role in millimeter/terahertz and even optical frequency bands. The performance of  waveguides determines the signal transmission efficiency of a whole system. The
	integrated  waveguide has an important influence in modern photonics\cite{01,02,03}. Recent years, the development of waveguides has made great progress.  In terms of materials, waveguides are from ordinary metal waveguides\cite{04}, dielectric optical fibers\cite{05,06}, and then to  some new waveguides based on new material like graphene\cite{07}, lithium niobate\cite{08}, perovskite\cite{09}, etc. Mechanistically, scientists have created sub-wavelength metal waveguides by harnessing the spoof surface plasmon polaritons (SSPPs)\cite{10,11}, a kind of method that change the plasma frequency by reduce the electron density of metal and other material. Also, they make coupled resonator optical waveguide (CROW) via imitating the atoms of the crystal lattice\cite{12}. Scientists made the Maxwell equation take on a form consistent with the Schrödinger equation by imposing periodic boundary condition, then explain the transmission phenomenon of photons by analogy with the transport of electrons in a solid crystal lattice, namely that photonic crystals\cite{13,14}.
	Surface-wave photonic crystal (SPC) waveguide\cite{15, 16, 17} demonstrates impressive performance based on SSPPs and PCs. SPC waveguides can introduce the defect mode into the bottom of photonic bandgap (PBG) by utilizing SSPPs to achieve deep-subwavelength waveguides. The transmission of electromagnetic waves (EMWs) in SPCs depends on the weak coupling between adjacent cavities from CROW theory. The closed SPC (CSPC) waveguides, based on metal-insulator-metal (MIM) waveguides\cite{18,19,20} and PCs, can introduce the waveguide mode outside the PBG of SPCs, exhibit deeper-subwavelength effect and multi-mode scenes. Obviously, CSPCs have great potential to integrate photonics. 
	\par In this paper, we conducted a comprehensive analysis of the performance of CSPC waveguides based on CROW theory, and greatly improved its transmission efficiency through structural redesign. At the same time, it was discovered that CSPC waveguide is a comprehensive physical model for explaining CROW theory.
	
	\begin{figure}[htb]
		\centering
		\includegraphics[width=0.45\textwidth]{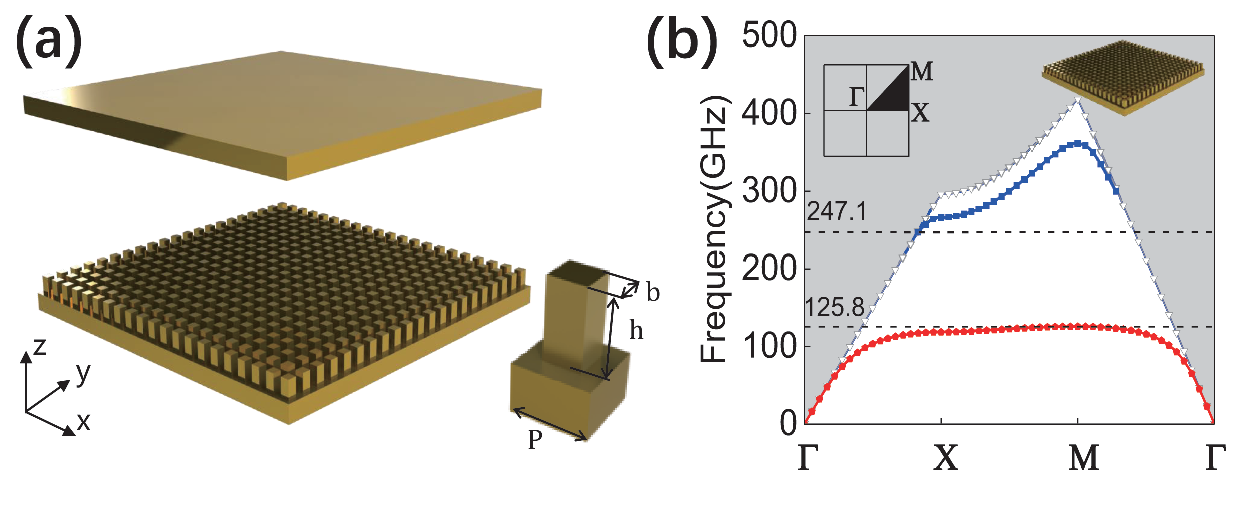}
		\caption{\label{fig:fig1}  (a) The 3D schematic diagram of CSPCs. The CSPCs consists of a top metal plate, a square array of square metallic rods and a bottom plate. (b) The dispersion relation of SPCs with the same size of CSPCs in (a).}
	\end{figure}
	\begin{figure*}[htb]
		\centering
		\includegraphics[width=\textwidth]{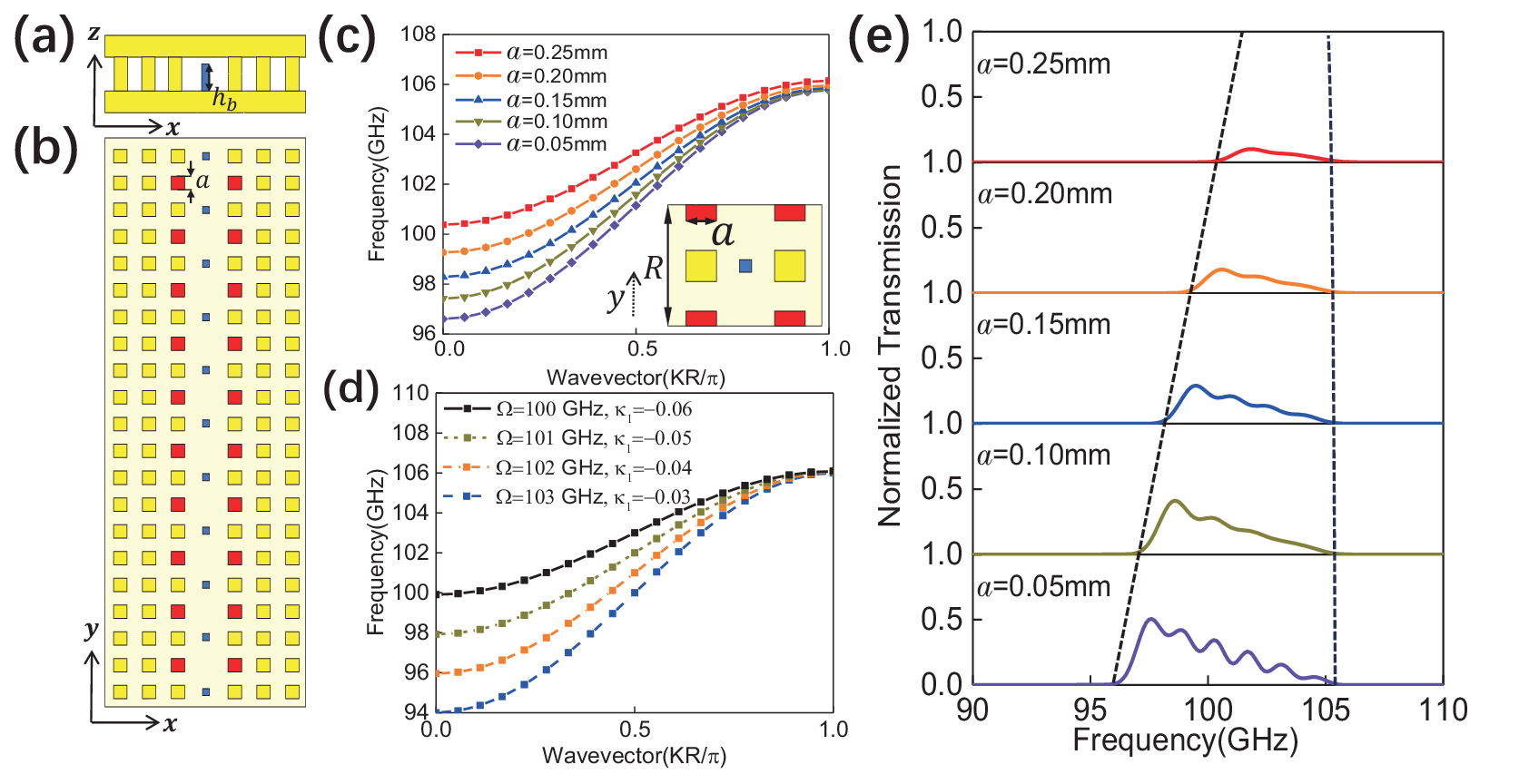}
		\caption{\label{fig:fig2}  (a) The $xz$-plane cross section of single-mode CSPC waveguides. (b) The $xy$-plane cross section of single-mode CSPC waveguides. (c) The dispersion relation of single-mode CSPC waveguides in (a) and (b) from commercial software with $a$  changed. (d) The dispersion relation of single-mode CSPC waveguides in (a) and (b) from CROW theory with $\Omega$ and $\kappa_1$ changed. (e) The normalized transmission of single-mode CSPC waveguides in (a) and (b) from commercial software with $a$ changed.}
	\end{figure*}
	\begin{figure*}[htb]
		\centering
		\includegraphics[width=\textwidth]{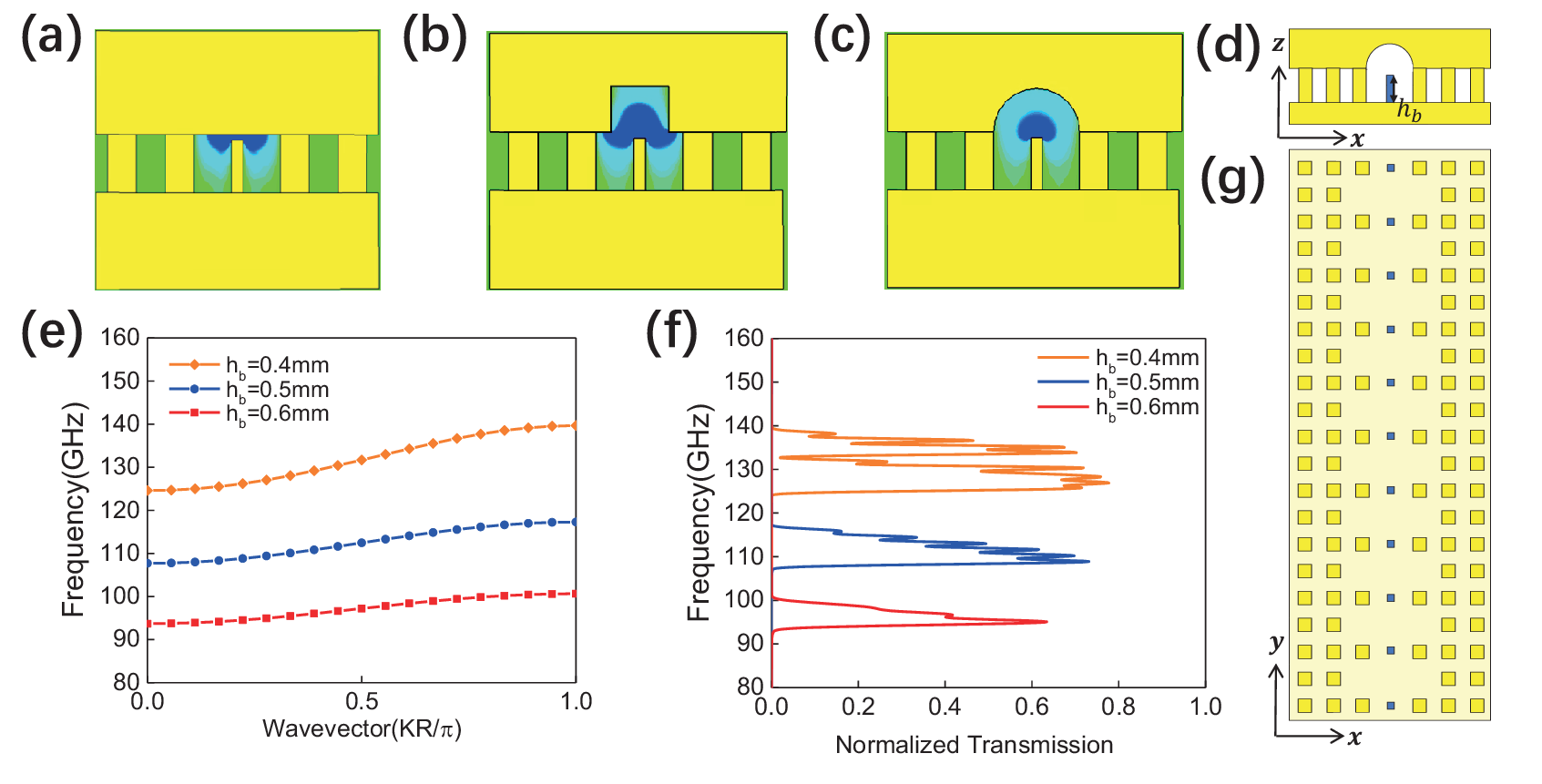}
		\caption{\label{fig:fig3}  (a) The Electric field cross-section of CSPC waveguides with a normal plate. (b) The electric field crossion-section of CSPC waveguides with rectangular cavity. (c) The electric field crossion-section of CSPC waveguides with semi-cylindercal cavity. (d) and (g) are the $xz$-plane $\&$ $xy$-plane of single-mode semi-cylindercal cavity CSPC waveguide. (e) The dispersion relation of waveguides in (d) and (g) with $\mathrm{h_b}$ changed. (f) The normalized transmission of waveguides in (d) and (g) with $\mathrm{h_b}$ changed.}
	\end{figure*}
	
	\section{Single-mode waveguides}
	CSPCs are sandwich structure, as shown in Fig. 1(a) consisting of the middle periodic metal rods and metal plate at the top and bottom. Its sizes are $P=0.5~\mathrm{mm}$, $h=0.5~\mathrm{mm}$, $b=0.25~\mathrm{mm}$. Without the top metal plate, it will be a tradition SPC structure. Fig. 1(b) is dispersion relation of SPC, the top left is the first Brillouin zone, the top right is the schematic diagram of SPCs. The grey curve is the dispersion relation of light, the red and blue curve are the first and second order dispersion curves of SPC. We can find there is a PBG from $125.8$ to $247.1~\mathrm{GHz}$.
	When we introduce defect rods into CSPCs, as the blue rods in Figs. 2(a) and 2(b), a waveguide mode can come out. Here, the sizes of blue defect rods are $h_b=0.45~\mathrm{mm}$, $d_b=0.1~\mathrm{mm}$.  The arrangement of defect rods is AOA…, O means no rod, A means one rod, as in Fig. 2(b). 
	Here, we calculate the dispersion relation by commercial software when the sidelength of red rods $a$ changes, shown in Fig. 2(c).The working frequency is still smaller than $125.8~\mathrm{GHz}$. The insert is a supercell of whole waveguides, whose period is $R=2*P$. As $a$ increases, there is a blue shift at the low frequency part in Fig. 2(c) and 2(e), and the high frequency part is not nearly influenced. According to CROW theory, the dispersion relation of single-mode CROW is
	\begin{equation}
		\omega^2_K=\Omega^2\frac{[1+{\sum_{n\neq 0}} \exp(-inKR)\beta_n]}{[1+\Delta\alpha+{\sum_{n\neq 0}} \exp(-inKR)\alpha_n]}
	\end{equation}
	When $\mathrm{n=1,-1}$, the dispersion relation of single-mode CROW is
	\begin{equation}
		\omega_K=\Omega[1-\Delta\alpha/2+\kappa_1\cos(KR)]
	\end{equation}
	Where, $K$ is wavevector in first Brillouin zone $-\pi/R\leq K\leq \pi/R$, $\Omega$ is the single-resonator mode frequency, $\Delta\alpha=\int d^{3}\mathbf{r}[\epsilon (\mathbf{r})-\epsilon_0 \mathbf{r})]\mathbf{E}_\Omega(\mathbf{r})\cdot \mathbf{E}_\Omega(\mathbf{r})$, the coupling factor
	$\kappa_1=\beta_1-\alpha_1=\int d^{3}\mathbf{r}[\epsilon_0 (\mathbf{r}-R\mathbf{e}_y)-\epsilon (\mathbf{r}-R\mathbf{e}_y)]\mathbf{E}_\Omega(\mathbf{r})\cdot \mathbf{E}_\Omega(\mathbf{r}-R\mathbf{e}_y)$, $\mathbf{E}_\Omega(\mathbf{r})$ is the high-Q modes of the individual resonators along a straight line parallel to the $\mathbf{e}_y$ axis, like in Fig. 2(c). Here, we assume $\Delta\alpha =0$\cite{12}. The insert of Fig. 2(c) can be seen as a unit resonator. When $a$ is changed and $R$ is stable, the inner space of unit resonator is bigger though the period of whole waveguide is the same. As we all know, for resonators, the bigger the inner space in unit cell, the lower their resonant frequency, so $\Omega$ will decrease. A large space will expand the integral range of $\kappa_1$, then the absolute value $|\kappa_1|$ will increase. When $\Omega$ and $|\kappa_1|$ change together, we can get the dispersion relation in Fig. 2(c). As a comparison, we select several data to get the analysis results based on Eq. (2), as in Fig. 2(d). The results by commercial software and CROW theory are consistent. 
	\par Fig. 2(e) is the normalized transmission from commercial software when $a$ changes. There is also a blue shift at the low frequency part, corresponding to the aforementioned results. The increment of coupling factor $|\kappa_1|$ naturally enhance coupling efficiency, then the transmission efficiency is also improved. As the result in Fig. 2(e), the normalized transmission get larger when $a$ is smaller. Another key point is bandwidth is larger with $\Omega$ decreasing and $|\kappa_1|$ increasing. For periodic structure, larger bandwidth must cause more serious oscillation in passband. It is easy to understand. We can obtain the group velocity $\nu_g(K)=d\omega_K/dK=-\Omega R \kappa_1 \mathrm{sin}(KR)$ by taking the derivative of Eq. (2). When $|\kappa_1|$ scale up, $\nu_g(K)$ is also larger. Naturally, Transmission effect will be enhanced and coupling effect will be attenuated\cite{21}. Thus, oscillation is most apparent at the point having the biggest group velocity, as shown in Fig. 2(e). However, the peak point is in the spot of a low frequency, and there is no high transmission at the position of high frequency. The reason is that a fixed length waveguide means longer electrical length $L/\lambda$ ($L$ is the length of waveguide) for higher-frequency EMWs. Hence, when the waveguide loss per length and length are stable, The higher the frequency, the greater the loss. So Fig. 2(e) shows that as the frequency of EMWs increases, the transmission efficiency gradually decreases.

	\par Generally, there are two parts of EMWs in waveguides: transmitting wave and evanescent wave. The basic idea of CROW theory is to guide wave by evanescent waves coupling between the two adjacent individual resonators. Hence, if the waveguide itself is directly in contact with the transmitting part of EMWs, the normalized transmission will be not good. So, we design three structures and simulate their electric field distribution in Figs. 3(a), 3(b) and 3(c). The difference is no defect in Fig. 3(a), rectangular defect cavity in Fig. 3(b) and semi-cylindrical defect cavity in Fig. 3(c) at the inside of top plate. We can find the waveguide contact the transmitting wave (the dark blue part) directly in Figs. 3(a) and 3(b). However, in Fig. 3(c), the waveguide is only connected to the evanescent wave (the light blue part). Then, based on the third structure as in Fig. 3(d) and 3(g), we calculate the dispersion relation and normalized transmission when the height of center defect rods changes. We can find the increment in $h_b$ will cause a whole blue shift on waveguide modes. From Fig. 3(f), we know the oscillation is very serious when $h_b=0.4~\mathrm{mm}<h$, and not apparent when $h_b=0.6~\mathrm{mm}>h$. Because EMWs are localized at the top of defect rods, they will be seriously influenced by the periodic side rods when  $h_b<h$. When $h_b>h$, the transmission part of EMWs does not need to directly contact the side rods, so oscillation nearly disappears. So, we improve the transmission efficiency from $10\%$ to $60\%$ and eliminate the inner oscillation after semi-cylindrical defect cavity introduced, red rods deleted and the height of defect $h_b=0.12*h$. 
	\begin{figure}[htb]
		\centering
		\includegraphics[width=0.45\textwidth]{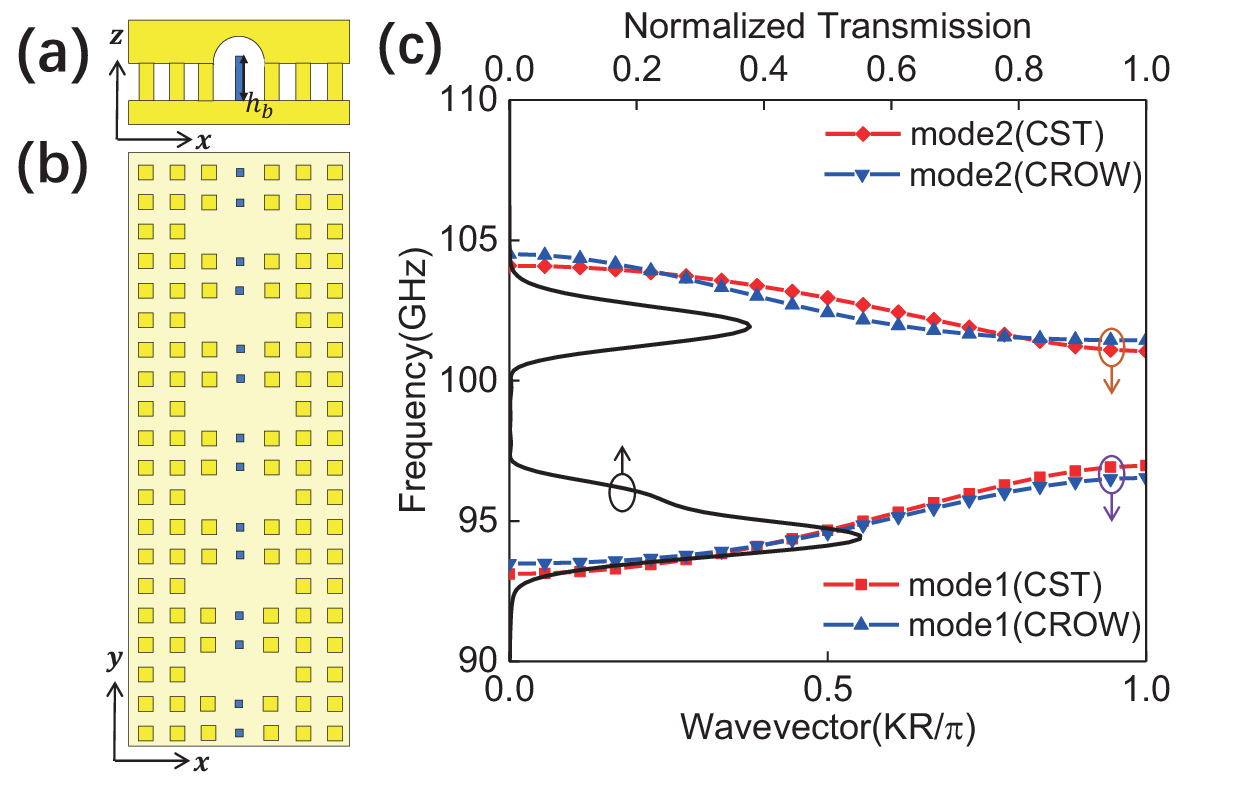}
		\caption{\label{fig:fig4}  (a) and (b) are the $xz$-plane $\&$ $xy$-plane of dual-mode semi-cylindercal cavity CSPC waveguide. (c) the dispersion relation  and  normalized transmission of waveguides in (a) and (b), the red dispersion curves are from commercial software, the blue dispersion curves are from CROW theory, all of them are corresponding to the bottom $x$ axis. The black is normalized transmission, connected with the top $x$ axis.}
	\end{figure}
	\begin{figure*}[htb]
		\centering
		\includegraphics[width=0.95\textwidth]{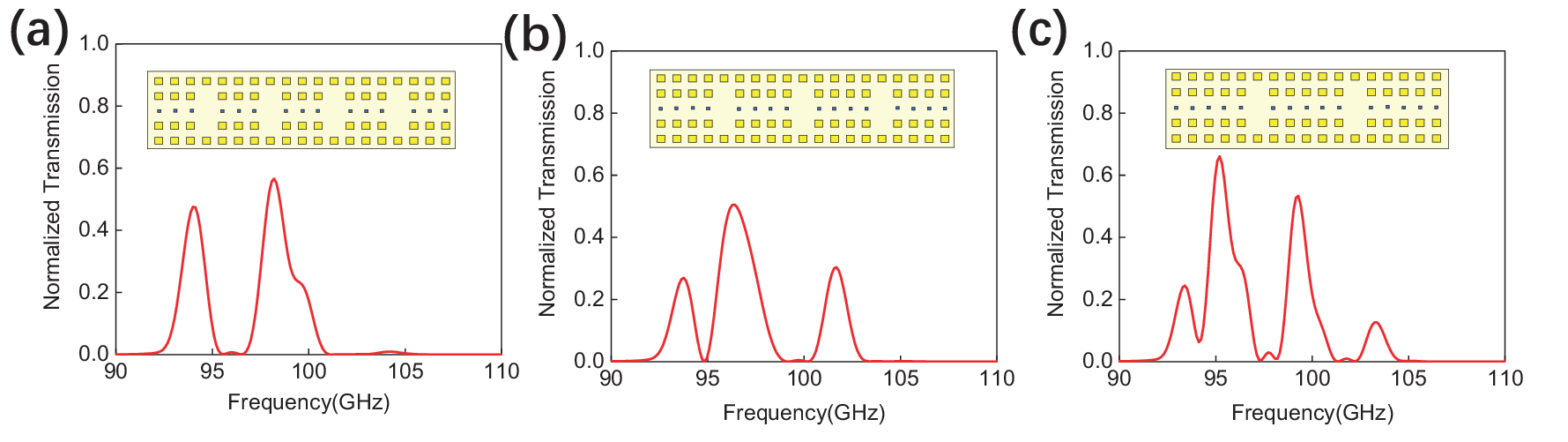}
		\caption{\label{fig:fig5} (a), (b) and (c) are the normalized transmission of 3,4,5-mode CSPC waveguides simulated. The insert pictures are their inner structures.}
	\end{figure*}

	\section{multi-mode waveguides}
	Based on these methods, we also calculate the performance of dual-mode and multi-mode waveguides. Dual-mode waveguides are shown in Fig. 4(a) and 4(b), $\mathrm{h_b}=0.6\mathrm{mm}<h$ here. Fig. 4(c) shows the dispersion relation and normalized transmission (black curve). The black curve is connected with the top $x$ axis. The red dispersion curves are calculated by commercial software, and the blue dispersion curves are by calculated CROW theory, all of them are corresponding to the bottom $x$ axis. From Figure 4(b), we know the arrangement of blue defect rods is AAOAAO…, so it should be $\mathrm{n=1,-2}$ or $\mathrm{n=2,-1}$ based on Eq. (1), we can get the dispersion function of dual-mode waveguide:
	\begin{subequations}
		\begin{equation}
			\omega^2_K=\Omega^2\frac{[1+\beta_{1} \exp(-iKR)+\beta_{-2} \exp(i2KR)]}{[1+\Delta\alpha+\alpha_{1} \exp(-iKR)+\alpha_{-2} \exp(i2KR)]}
		\end{equation}
		
		\begin{equation}
			\omega^2_K=\Omega^2\frac{[1+\beta_{2} \exp(-i2KR)+\beta_{-1} \exp(iKR)]}{[1+\Delta\alpha+\alpha_{2} \exp(-i2KR)+\alpha_{-1} \exp(iKR)]}
		\end{equation}
	\end{subequations}
	Here, $\Omega=98.5\mathrm{GHz}$, $\Delta\alpha=0.08$, $\beta_1=-\beta_{-1}=0.004$, $\beta_2=-\beta_{-2}=0.003$,$\alpha_1=-\alpha_{-1}=0.008$,$\alpha_2=-\alpha_{-2}=0.03$.
	\par  It is obvious that normalized transmission matches the numerical and analytical dispersion relation very well. Compared with our previous work\cite{22}, transmission is improved from about $10\%$ to $50\%$ (mode 1) and $40\%$ (mode 2). Multi-mode waveguides are basically in the same situation, but a special phenomenon is there are only $\mathrm{n}-1$ passband in $n$-mode waveguides, shown in Fig. 5. As shown in Figs. 2(c) and 2(d), there exist a maximum value $106~\mathrm{GHz}$ for the frequency of EMWs coupled. And the position that highest modes disappear is exact $106~\mathrm{GHz}$ in three cases of Fig. 5. We can get that the highest forbidden bands are located at the point of $106~\mathrm{GHz}$. In this case, $n$-mode waveguides only have $\mathrm{n}-1$ passbands.

	\section{Conclusion}
	In conclusion, we analyze the performance of CSPC waveguides based on coupled-resonator optical waveguide theory. In the case of single mode, firstly, we improve its transmission efficiency via increasing coupling factor $|\kappa_1|$ by shrinking parameter $a$, but $|\kappa_1|$ enhanced can also bring serious oscillation since transmission effect get bigger. Secondly, we design semi-cylindrical defect cavity to elimitate the oscillation by stop the waveguide itself from being in contact with the transmission part of EMWs. Thirdly, through analysis to the height of defect rods  $h_b$, we improve the transmission from $\mathrm{10 \%}$ to $\mathrm{60 \%}$ and eliminate the oscillation at the same time. And the method is also proved to be feasible in the multi-mode scenario. And it is obvious that CROW theory is proved properly through CSPC waveguide model.

	~\\
	\textbf{Fundings.} This work was supported by National Natural Science Foundation of China (Grant Nos. 12333012, 61927813, 61975225) and Science and Technology Commission of Shanghai Municipality (21DZ1101102).

	~\\
	\textbf{Disclosures.} The authors declare no conflicts of interest.

	~\\
	\textbf{Data availability.} Data underlying the results presented in this paper are not publicly available at this time but may be obtained from the authors upon reasonable request.

	\bibliography{reference}

\end{document}